\RequirePackage{amsmath}
\documentclass[smallextended]{svjour3}
\usepackage{graphics}
\usepackage{amsmath,amssymb}
\usepackage{natbib}
\usepackage{bm}
\usepackage{xcolor}
\usepackage{hyperref}
\usepackage[title]{appendix}

\smartqed
\journalname{Exp Astron}

\begin{document}
\sloppy

\title{Attitude determination for nano-satellites -- II. Dead reckoning with a multiplicative extended Kalman filter}

\author{J\'anos Tak\'atsy \and Tam\'as Boz\'oki \and \\ Gergely D\'alya \and Korn\'el Kap\'as \and \\ L\'aszl\'o M\'esz\'aros \and Andr\'as P\'al}

\institute{%
J. Tak\'atsy, T. Boz\'oki, G. D\'alya, K. Kap\'as, L. M\'esz\'aros, A. P\'al
        \at Konkoly Observatory of the Research Centre for Astronomy and Earth Sciences, Budapest, Hungary \\
E-mail: takatsy.janos@wigner.hu \\
J. Tak\'atsy, G. D\'alya, K. Kap\'as, L. M\'esz\'aros, A. P\'al
        \at E\"otv\"os Lor\'and University, P\'azm\'any P\'eter stny. 1/A, Budapest H-1117, Hungary\\
J. Tak\'atsy
        \at Institute for Particle and Nuclear Physics, Wigner Research Centre for Physics, Konkoly-Thege Mikl\'os \'ut 29-33, Budapest H-1121, Hungary \\
T. Boz\'oki
        \at Institute of Earth Physics and Space Science (ELKH EPSS), Csatkai Endre utca 6-8, Sopron H-9400, Hungary \\
T. Boz\'oki
        \at Doctoral School of Environmental Sciences, University of Szeged, Aradi v\'ertan\'uk tere 1, Szeged H-6720, Hungary
}

\maketitle

\begin{abstract}
This paper is the second part of a series of studies discussing a novel attitude determination method for nano-satellites. Our approach is based on the utilization of thermal imaging sensors to determine the direction of the Sun and the nadir with respect to the satellite with sub-degree accuracy. The proposed method is planned to be applied during the Cubesats Applied for MEasuring and LOcalising Transients (CAMELOT) mission aimed at detecting and localizing gamma-ray bursts with an efficiency and accuracy comparable to large gamma-ray space observatories. In our previous work we determined the spherical projection function of the MLX90640 infrasensors planned to be used for this purpose. We showed that with the known projection function the direction of the Sun can be located with an overall accuracy of $\sim40^\prime$.

In this paper we introduce a simulation model aimed at testing the applicability of our attitude determination approach. Its first part simulates the orbit and rotation of a satellite with arbitrary initial conditions while its second part applies our attitude determination algorithm which is based on a multiplicative extended Kalman filter. The simulated satellite is assumed to be equipped with a GPS system, MEMS gyroscopes and the infrasensors. These instruments provide the required data input for the Kalman filter. We demonstrate the applicability of our attitude determination algorithm by simulating the motion of a nano-satellite on Low Earth Orbit. Our results show that the attitude determination may have a 1$\sigma$ error of $\sim30'$ even with a large gyroscope drift during the orbital periods when the infrasensors provide both the direction of the Sun and the Earth (the nadir). This accuracy is an improvement on the point source detection accuracy of the infrasensors. However, the attitude determination error can get as high as 25$^{\circ}$ during periods when the Sun is occulted by the Earth. We show that following an occultation period the attitude information is immediately recovered by the Kalman filter once the Sun is observed again.

\keywords{Space vehicle instruments (1548), Stellar tracking devices (1633), Pointing accuracy (1271), Astrometry (80)}
\end{abstract}

\section{Introduction}
\label{sec:introduction}
Owing to the enormous funding requirements, satellite missions were only conducted by the economically most powerful countries of the world in the first few decades of the space era. However, as a consequence of the explosive technological development, small satellite missions with substantially lower funding requirement -- for instance, nano-satellites including as CubeSats -- became a viable alternative for traditional large-size and high-cost satellites, which made space an achievable goal also for countries/organizations with less financial resources. The last few decades brought along a lot of such missions with more and more scientific aims being targeted by them.

One of the technological difficulties that needs to be handled in connection with small satellite missions is the accurate determination of the satellite's actual orientation, i.e. its attitude. While on large-size satellites this information is usually provided by costly, large-size star trackers, which determine the attitude based on the angular distribution of bright stars in their field of view, these systems do not fit the very restricted size and power budget criteria of nano-satellites. In our recent paper \citep{kapas2021} we proposed a new, cost-efficient approach to this problem which is based on the utilization of thermal imaging infrasensors. For this purpose we chose the MLX90640 infrasensor of \cite{MLX}, which is a small-size, low-cost sensor having 32$\times$24 pixels and a relatively large, 110$\times$75 degree field of view. This coverage by a single sensor implies that six of these sensors, placed on the six sides of a cube, could cover the full sphere, see Figure~4 of \cite{kapas2021}. This technology might be suitable to even smaller satellites in similar missions like GRBAlpha \citep{pal2020}. As the spherical projection function of MLX90640 infrasensors (to be used for this purpose) is now known with an overall accuracy of $\sim40^\prime$ \citep{kapas2021} we now turn to the next step and introduce a simulation model for testing the applicability of our attitude determination approach. As we outline in our recent paper \citep{dalya2020}, our method is based on a multiplicative extended Kalman filter that uses the information provided by the infrasensors (direction of the Sun and the nadir in the satellite's coordinate frame), the GPS system (the location of the satellite in the Earth centered coordinate frame) and the MEMS gyroscopes (angular velocity of the satellite) carried by the satellite.

The layout of the paper is the following. In Section~\ref{sec:Simulation} we describe the simulation of the satellite dynamics and introduce the multiplicative extended Kalman filter method. We demonstrate our results in Section~\ref{sec:results}, and we summarize our conclusions in Section~\ref{sec:conclusions}. A detailed description of the equations used for the Kalman filter can be found in Appendix~\ref{sec:appendix}.

\section{Simulation model for testing the on-board attitude determination algorithm}
\label{sec:Simulation}
The attitude determination algorithm we developed is aimed to run on-board and therefore it needs to be tested for the different situations possible during a space mission. For this purpose we built a simulation model where all parts of the attitude determination process can be tested independently and as a whole as well. The first part of the code simulates the dynamics of the Sun-Earth-satellite system while its second part determines the attitude of the satellite by applying a multiplicative extended Kalman filter to the simulated data provided by the first part of the code. The goal of this process is to see how the recovered attitudes compare to the 'real' ones.

\subsection{Simulation of the satellite dynamics}
\label{ssec:SatDyn}

\begin{figure}
\begin{center}
    \resizebox{10cm}{!}{\includegraphics{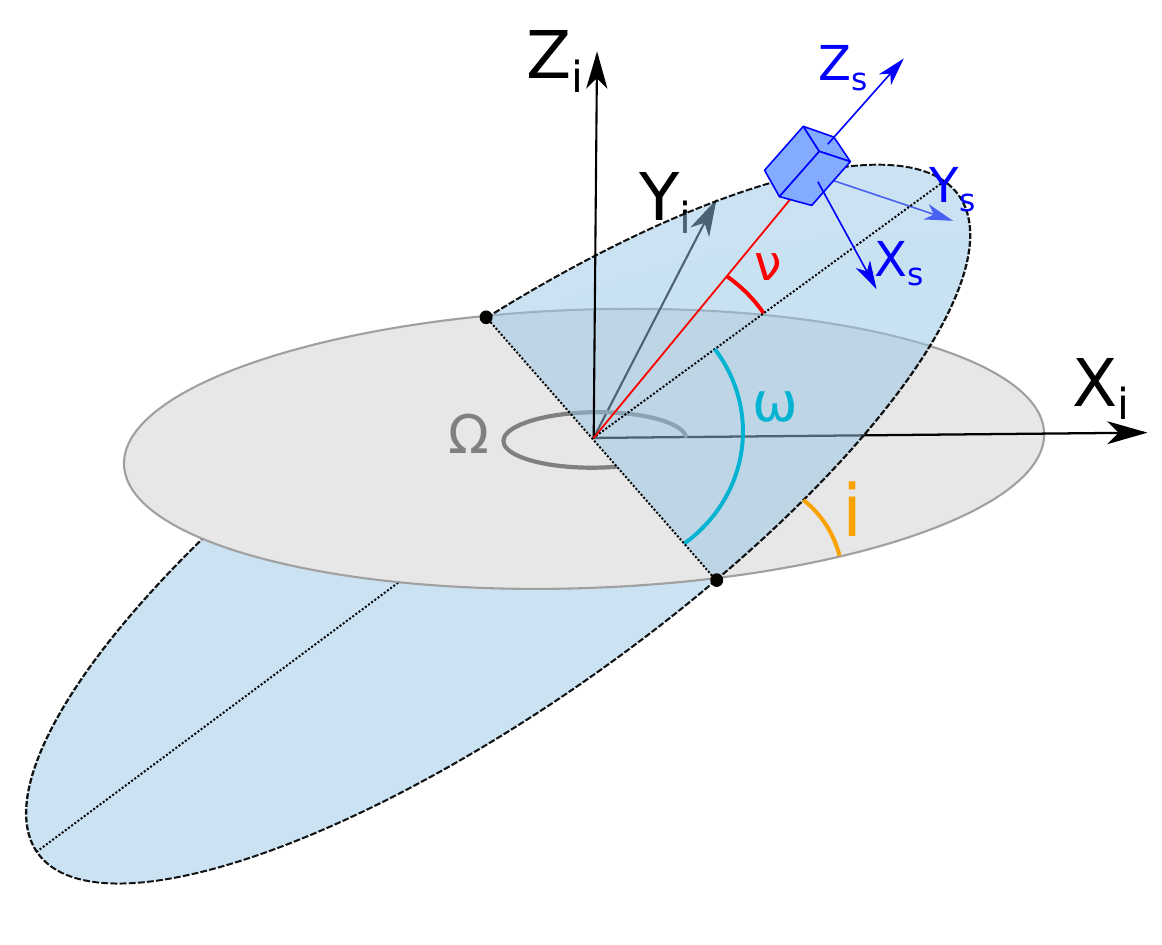}}
    \caption{The two main coordinate systems used in the simulations. Index $i$ denotes the J2000 system with the Earth in the origin. The $Z_i$ axis coincides with the rotation axis of the Earth in January 1, 2000 at 12:00 TT while the $X_i$ axis points to the vernal equinox at the same epoch. The axes of the satellite's coordinate frame are denoted by index $s$. The orbital parameters displayed in the figure are $\Omega$ (longitude of ascending node), $i$ (inclination), $\omega$ (argument of periapsis) and $\nu$ (true anomaly).}
    \label{fig:cosys}
    \end{center}
\end{figure}

This part of the code calculates the position and the attitude of the satellite in the Earth centered J2000 reference system (where the $X$ and $Z$ axes point towards the former positions of the vernal equinox and the Earth's rotation axis in January 1, 2000 at 12:00 TT, respectively) as well as the position of the Sun and the Earth in the satellite's coordinate system (where the origin of the system is fixed to the center of mass of the cubesat and the axes are parallel to its edges, see Fig. \ref{fig:cosys}). In the code time is expressed in Julian dates and the GPS to J2000 coordinate transformations are implemented as well. The code also determines the 'night' part of the orbit, i.e. where the Sun is occulted by the Earth.

The orbit of a satellite can be characterized by five orbital elements ($\Omega$, longitude of ascending node; $i$, inclination; $\omega$,  argument of periapsis; $e$, eccentricity; $h$, altitude of satellite orbit at perigeum; see Fig \ref{fig:cosys}.) which remain constant when assuming a spherical Earth. The actual position of the satellite on this elliptical orbit is given via the true anomaly ($\nu$) which is obtained from the Kepler equation. However, the oblateness of the Earth introduces perturbations, from which the $J_2$ perturbation has the largest magnitude, and therefore it has to be taken into account while simulating the satellite dynamics. The main effect of the $J_2$ perturbation is on $\Omega$ and $\omega$. The time derivatives of these orbital elements are the following \citep{kozai1959}:
\begin{equation}
    \dot{\omega}=\frac{3}{4} n J_2 \left( \frac{R_e}{p} \right)^2 (5\cos ^2 i -1 )
\end{equation}
\begin{equation}
    \dot{\Omega} = -\frac{3}{2} n J_2 \left( \frac{R_e}{p} \right)^2 \cos i
\end{equation}
where $R_e$ is the radius of the Earth, $p=a(1-e^2)$, $n=\sqrt{GM_e / a^3}$ and $J_2 \simeq 1.082629 \cdot 10^{-3}$.

Next we discuss the description of the rotation of the satellite. As the variation of the gravitational field is negligible in the range of the satellite's dimensions and we did not apply any kind of attitude control in our model, the net angular momentum transfer due to external torques is negligible during one orbit. We also assumed that the satellite frame coincides with the principal frame, however, in case it does not, an additional constant rotation needs to be taken into account between the two frames. The time evolution of the attitude then can be determined using Euler's rotation equation, which describes the evolution of the angular velocity of a rotating rigid body represented in its principal frame \citep[see e.g.][]{coutsias2004}:
\begin{equation}
\label{eq:euler}
    \frac{\mathrm{d}}{\mathrm{d}t}\begin{pmatrix}
    \omega_1 \\
    \omega_2 \\
    \omega_3
    \end{pmatrix}=
    \begin{pmatrix}
    \dfrac{I_2-I_3}{I_1}\omega_2\omega_3 \\
    \dfrac{I_3-I_1}{I_2}\omega_3\omega_1 \\
    \dfrac{I_1-I_2}{I_3}\omega_1\omega_2
    \end{pmatrix} .
\end{equation}
where $\bm{\omega}$ is the angular velocity vector of the satellite represented in the principal frame (which coincides with the satellite frame), and $I_1$, $I_2$ and $I_3$ are the moments of inertia corresponding to the $x$, $y$ and $z$ axes, respectively. Note that Eq.~(\ref{eq:euler}) could imply complex motions, such as tumbling\footnote{see \url{https://youtu.be/1n-HMSCDYtM}}, which we can readily reproduce within our model (see supplementary material).

The attitude can then be calculated by an additional integration over the angular velocities. In this work we use unit quaternions to represent the attitude of the satellite as $\mathbf{q}^s=[\mathbf{n}\sin(\gamma/2),\cos(\gamma/2)]^{\mathrm{T}}$, where $\mathbf{n}$ is the axis of the rotation that transforms the J2000 coordinate frame to the satellite frame, $\gamma$ is the rotation angle and the superscript $\mathrm{T}$ denotes transposition (quantites represented in the satellite frame are denoted by the superscript 's'). The quaternion kinematics is given by the following equation \citep[see e.g.,][]{crassidis2012,baroni2018}:
\begin{equation}
    \label{eq:qdin}
    \dot{\mathbf{q}}^s = \frac{1}{2}\mathbf{\Omega}\left( \bm{\omega}^s \right) \mathbf{q}^s ,
\end{equation}
with the $4\times 4$ matrix
\begin{equation}
    \mathbf{\Omega}\left( \bm{\omega}^s \right) = 
    \begin{bmatrix}
    -\mathbf{S}(\bm{\omega}^s) & \bm{\omega}^s \\
    -(\bm{\omega}^s)^{\mathrm{T}} & 0 
    \end{bmatrix} ,
\end{equation}
where $\mathbf{S}(\bm{\omega}^s)$ is the matrix representation of the cross product ($\bm{\omega}^s\times)$:
\begin{equation}
    \mathbf{S}(\bm{\omega})=
    \begin{bmatrix}
    0 & -\omega_3 & \omega_2 \\
    \omega_3 & 0 & -\omega_1 \\
    -\omega_2 & \omega_1 & 0
    \end{bmatrix} .
\end{equation}

\subsection{Attitude determination with Kalman filter}
\label{ssec:Kalman}

The Kalman filter is an algorithm that provides an efficient way to estimate the state of a dynamic system by a series of measurements with inaccuracies over time \citep{kalman1960}. The estimates produced by the algorithm are more accurate than those based on a single measurement alone since the joint probability distribution of the variables is estimated for each discrete time-step of the process. This leads to the minimization of the mean of the squared error of the estimates. 

The Kalman filter works in a two-step process with a prediction step (time update) and a measurement step. In the prediction step the filter propagates the estimates of state and uncertainties from current to the next time step. In the measurement step the state of the system is measured with some error and the estimate is updated by the weighted average of the estimate and the measurement, where the weights are determined by the respective uncertainties.

In our specific objective the Kalman filter serves to combine the infrasensor data with the angular velocity information provided by the MEMS gyroscopes. Although the 3 axis MEMS gyroscopes yield an accurate attitude information on a short time period, due to the error accumulation effect known as gyroscope drift an absolute attitude information is required as well, which is provided by the thermal imaging infrasensors in our case. In earlier works this kind of absolute attitude information was usually gathered by a 3-axis magnetometer and an optical Sun sensor (see e.g., \citealt{ni2011, springmann2012, baroni2018, gaber2020}). The use of infrasensors is more convenient in the sense that as opposed to magnetometers they can be built-in parts of the satellite and do not need an external boom to be mounted on. The infrasensors determine the direction of the Sun and the nadir in the satellite's coordinate frame, which vectors are known in the Earth centered reference frame as well owing to the location information provided by the onboard GPS. The rotation that transforms the reference frame to the satellite frame (which is indeed equivalent to the attitude of the satellite) is unequivocally determined by the pair of these two vectors in the two coordinate frames. Hence we get a prediction of the system's state from the gyroscope which can be corrected by the absolute attitude information provided by the infrasensors.

Since the quaternion kinematics, described by Eq. (\ref{eq:qdin}), is nonlinear in the variables $\omega^s$ and $\mathbf{q}^s$ the utilization of an extended Kalman filter is necessary. We use the Multiplicative Extended Kalman Filter (MEKF) method \citep{lefferts1982}, where a multiplicative error state $\delta\mathbf{q}$ is introduced, which represents a small rotation from the predicted attitude -- which contains measurement errors -- to the actual attitude (from now on we omit superscripts, since everything is understood to be represented in the satellite frame, unless noted otherwise)\footnote{Note that the quaternion product $\otimes$ is conventionally defined here such that $ijk=1$ instead of the more commonly used $ijk=-1$}:
\begin{equation}
\label{eq:dq+-}
    \delta\mathbf{q} = \mathbf{q} \otimes \hat{\mathbf{q}}^{-1} ,
\end{equation}
where the circumflex '$^\wedge$' denotes the expected (or predicted) value of a quantity. With this multiplicative approach the conservation of the unit quaternion length is guaranteed and the problem of singular covariance matrices, encountered in the additive approach, is avoided as well.

To describe gyroscope measurements we use the model of \cite{lefferts1982}, where in addition to a zero mean Gaussian error $\bm{\eta_\omega}$ a time dependent bias vector $\bm{\beta}$ is also introduced, the motion of which is determined by a random walk. Hence, the measured angular velocity $\bm{\omega}_m$ is given by
\begin{equation}
    \bm{\omega}_m = \bm{\omega} + \bm{\beta} + \bm{\eta_\omega} ,
\end{equation}
\begin{equation}
    \dot{\bm{\beta}} = \bm{\eta_\beta} ,
\end{equation}
where the Gaussian processes $\bm{\eta_\omega}$ and $\bm{\eta_\beta}$ have covariance matrices $\sigma_\omega^2 \mathbf{I}_{3\times3}$ and $\sigma_\beta^2 \mathbf{I}_{3\times3}$, respectively. Therefore the estimated value of the angular momentum is
\begin{equation}
    \hat{\bm{\omega}} = \bm{\omega}_m - \hat{\bm{\beta}} .
\end{equation}
The state space model is then given by the equations:
\begin{equation}
    \dot{\mathbf{q}} = \frac{1}{2}\mathbf{\Omega}\left( \hat{\bm{\omega}} \right) \mathbf{q} ,
\end{equation}
\begin{equation}
    \dot{\bm{\beta}} = 0 .
\end{equation}

The predicted values of the angular momentum and bias vectors are updated through the Kalman filter using the position of the Sun and the nadir both as seen by the satellite and as calculated in the inertial frame. A vector in the inertial frame is transformed to the satellite frame by
\begin{equation}
    \mathbf{r}^s = \mathbf{A}(\mathbf{q}^s) \mathbf{r}^i .
\end{equation}
It can then be shown that a small multiplicative quaternion error $\delta\mathbf{q}$ creates a small $\delta \mathbf{r}^s$ deviation detemined by the following equation \citep{crassidis2012}:
\begin{equation}
\label{eq:dq}
    \delta \mathbf{r}^s = 2\mathbf{S}\left( \mathbf{A}\left( \hat{\mathbf{q}}^s \right) \mathbf{r}^i \right) \delta \mathbf{q}_3 ,
\end{equation}
where $\delta\mathbf{q}_3$ is a three component vector containing the imaginary part of the multiplicative error state $\delta\mathbf{q}$. This determines the so called sensitivity matrix, which is then used by the Kalman filter to calculate the multiplicative error state from the $\delta \mathbf{r}^s$ quantities.

Since measurements are made at discrete time-steps, to implement these equations one must first discretize the kinematical equations. The Kalman filter then can be applied after each time-step to refine the attitude information predicted by the kinematical equations. The detailed discretized equations can be found in Appendix~\ref{sec:appendix}.

\section{Simulation results}
\label{sec:results}

\begin{figure}
\begin{center}
    \resizebox{12cm}{!}{\includegraphics{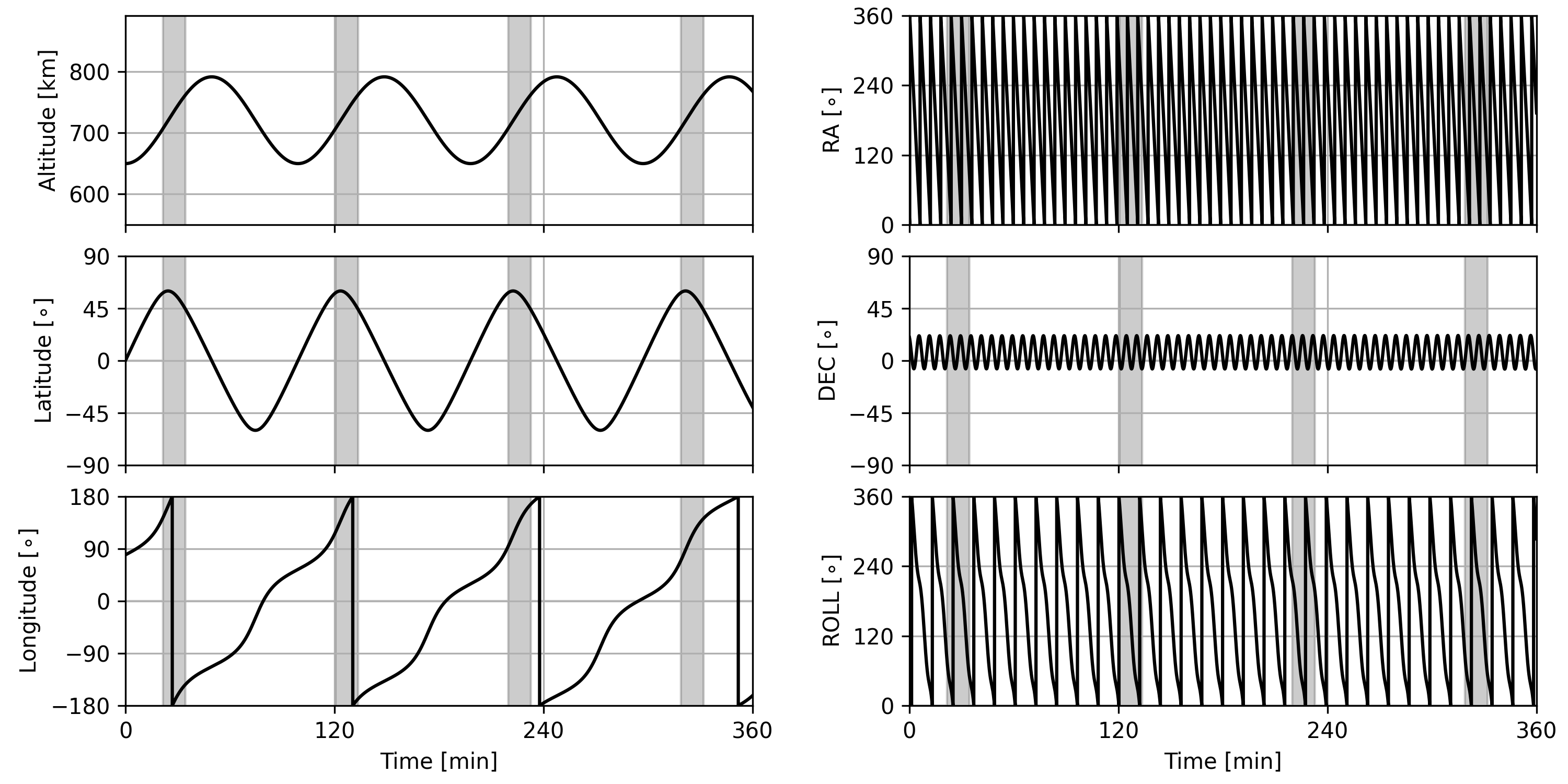}}
    \caption{The orbit and rotation of the simulated satellite. Its position is described by its altitude, latitude and longitude (left) while its attitude is described by its right ascension ($\alpha$), declination ($\delta$) and roll ($\rho$) (right). For the different parameters that characterize the satellite's motion we refer to the main text. Shaded areas represent those parts of the orbit where the Sun is occulted by the Earth.}
    \label{fig:fig2}
    \end{center}
\end{figure}

The applicability of our attitude determination approach is demonstrated by simulating a satellite on Low Earth Orbit (LEO) described by the following parameters:
\begin{itemize}
    \item $\Omega = 0^\circ$, $i = 60^\circ$, $\omega = 0^\circ$, $e = 0.01$, $h = 650$ km,
    \item $I_1 = 2.75 \times 10^{-4}$ kg m$^2$, $I_2 = 2.75 \times 10^{-4}$ kg m$^2$, $I_2 = 5.5 \times 10^{-5}$ kg\ m$^2$, 
    \item $L_0 = [-4.4 \times 10^{-6},\ 1.925 \times 10^{-6},\ -6.05 \times 10^{-7}]$ kg m$^2$/s,
\end{itemize}
where $L_0$ denotes the initial angular momentum of the satellite in the satellite frame. The chosen $I_1$, $I_2$ and $I_3$ values correspond to a 3U CubeSat with a size of $10$x$10$x$30$ cm and total mass of $3.3$ kg. The initial attitude was selected randomly. Figure \ref{fig:fig2} shows how the position and the attitude of the satellite changes during 6 hours on such an orbit. The attitude is represented in the form of right ascension ($\alpha$), declination ($\delta$) and roll ($\rho$). The conversion rule between these angles and the quaternion representation of the attitude is given by the following formulas:
\begin{align}
    \alpha &= \arg(q_1 q_3 + q_2 q_4, q_2 q_3 - q_1 q_4) , \nonumber \\
    \delta &= \arg(q_4^2 + q_3^2 - q_2^2 - q_1^2, 2\sqrt{(q_1^2+q_2^2)(q_3^2+q_4^2)}) , \nonumber \\
    \rho &= \arg(q_1 q_3 - q_2 q_4, -q_2 q_3 - q_1 q_4) ,
\end{align}
where $\arg(x,y)$ gives the $\varphi$ phase factor of the complex number $x+iy=r\cdot e^{i\varphi}$. Note also that the domain of $\delta$ is $[0^\circ,90^\circ]$, while it is $[0^\circ,360^\circ)$ for $\alpha$ and $\rho$. In the figure shaded areas represent those parts of the orbit where the Sun is occulted by the Earth, i.e. where the number of measured vectors for the MEKF is reduced to one.

Since MEMS gyroscopes are available with various precision we investigated three different cases for attitude determination characterized by three different values for gyroscope drifts. For the largest error case we used $\sigma_{\omega} = 4.89 \times 10^{-3}$ rad/s$^{1/2}$ and $\sigma_{\beta} = 3.14 \times 10^{-4}$ rad/s$^{3/2}$ as proposed by \cite{baroni2018}, while for our standard and low-error case we used errors $0.1$ and $0.3$ times those of the high-error case, respectively. The initial parameters of the MEKF were chosen as follows:
\begin{itemize}
    \item $\beta_0 = [0,0,0]$,
    \item $P_0 = \mathrm{diag}([0.25,0.25,0.25,0.01,0.01,0.01])$.
\end{itemize} 
$q_0$ was selected randomly and the standard deviation of the measured vectors had been set to $0.012$ rad ($\sim 40'$, in accordance with our previous result on the pointing accuracy of MLX90640 infrasensors) and had been added to the input vectors. Sensor data were sampled at $1$ Hz. 

\begin{figure}
\begin{center}
    \resizebox{12cm}{!}{\includegraphics{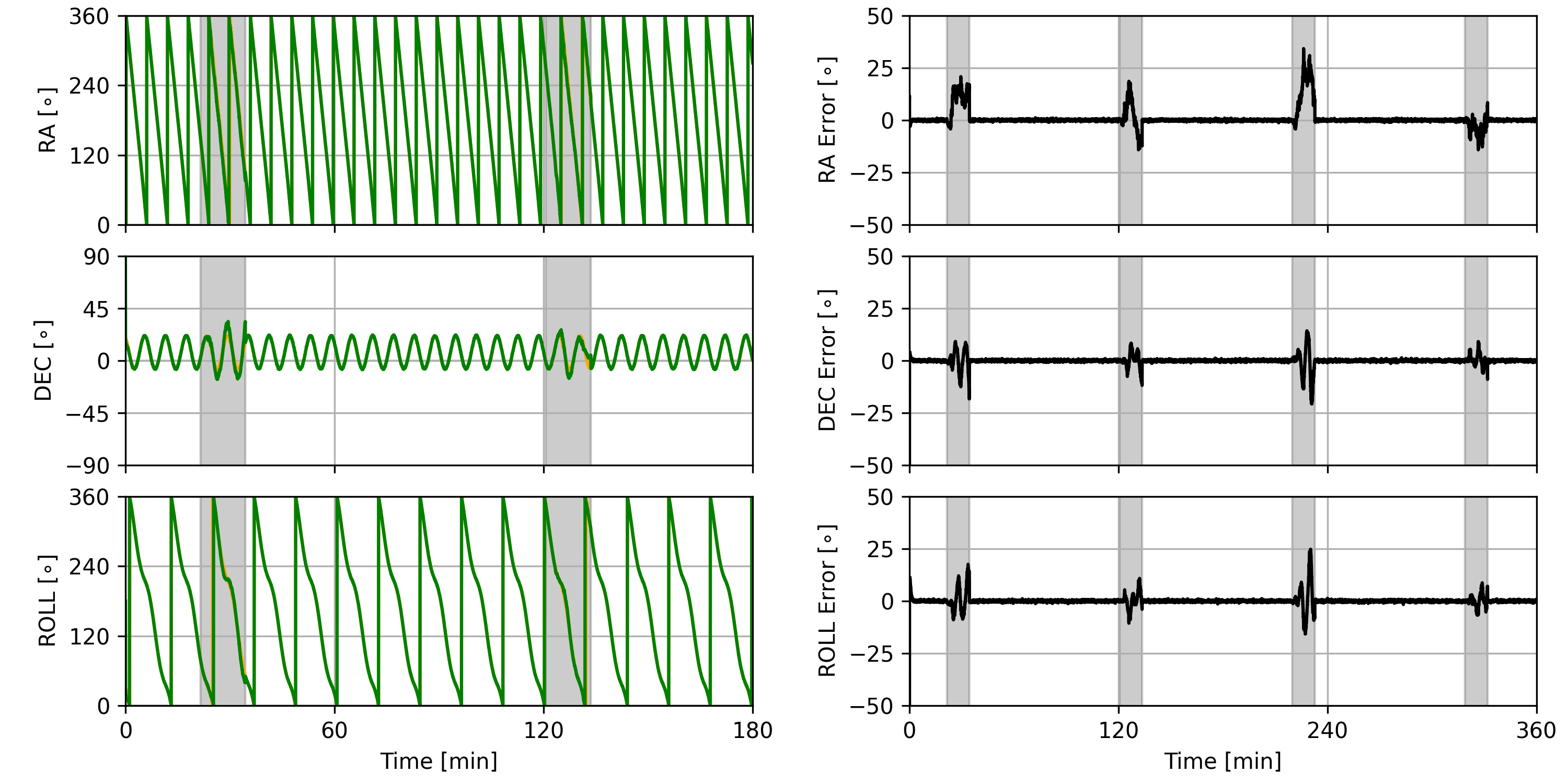}}
    \caption{The real (orange) and the MEKF recovered (green) attitude for the simulation shown in Fig. \ref{fig:fig2} for our standard choice of gyroscope error (left), and the error of the attitude determination, i.e. the difference between the real and the recovered attitude elements for the same orbital configuration (right). Shaded areas represent parts of the orbit where the Sun is occulted by the Earth.}
    \label{fig:fig3}
    \end{center}
\end{figure}

The recovery of the attitude on the orbit shown in Fig. \ref{fig:fig2} for our standard choice of gyroscope error is presented in Fig.~\ref{fig:fig3}. The left panels of Fig. \ref{fig:fig3} show that the attitude elements are well recovered when the MEKF works with two input vectors ('day'), i.e. when the infrasensors provide both the direction of the Sun and the Earth (the nadir), while the accuracy breaks down significantly when there is only one input vector (only the nadir direction) available for the MEKF ('night'). This behavior is not surprising, since we are lacking the minimum of two linearly independent vectors necessary to gain information about the absolute attitude of the satellite, and since the bias instability of MEMS gyroscopes is relatively high. The right panels of Fig. \ref{fig:fig3} show that the difference between the real and the recovered attitude elements may reach 25$^{\circ}$ during the 'night' phase in our standard case.

\begin{figure}
\begin{center}
    \resizebox{12cm}{!}{\includegraphics{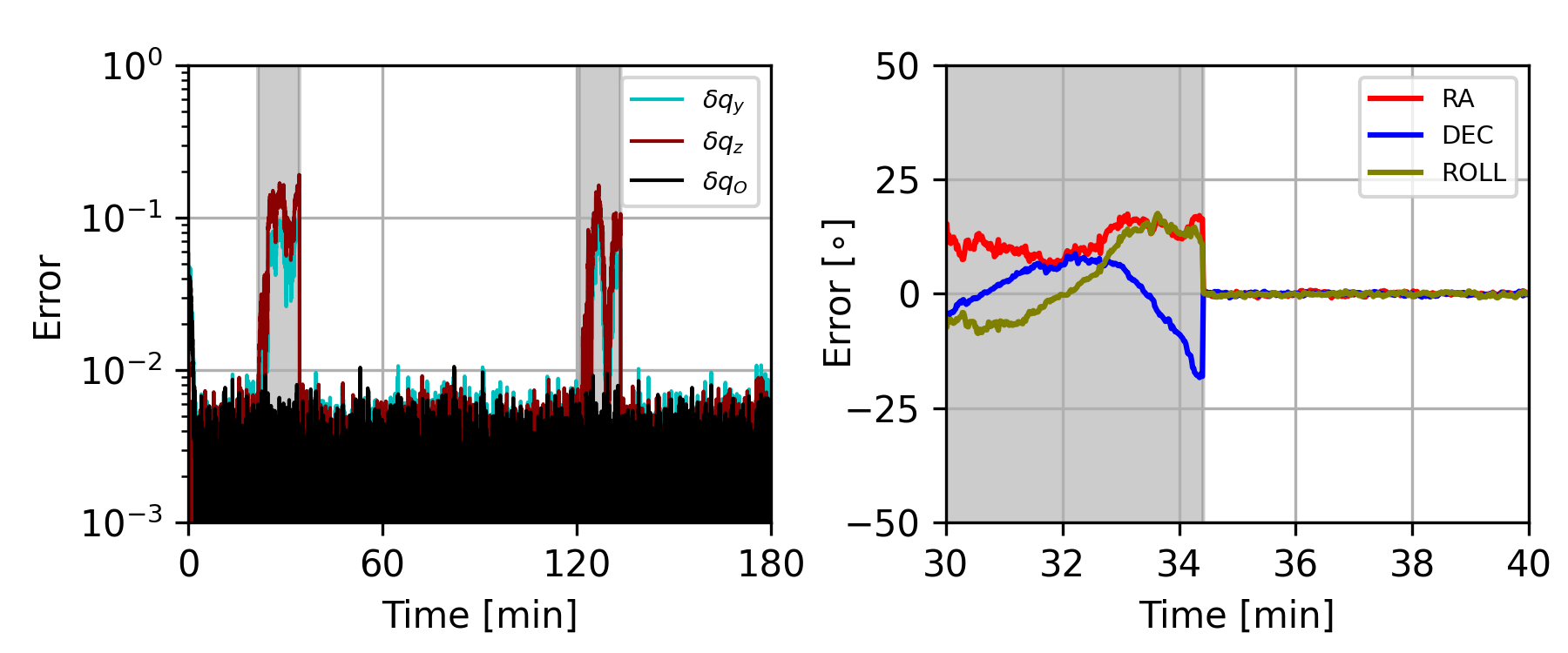}}
    \caption{The $y$ and $z$ components of the quaternion error states, as well as the component corresponding to the rotation in the satellite's orbital plane (left), and the errors of the attitude elements around a 'night' to 'day' transition (right).}
    \label{fig:fig4}
    \end{center}
\end{figure}

Even though the accuracy of recovering the independent attitude parameters breaks down during 'night', the errors are correlated even in this case due to the information gained from observing the horizon. This is shown in the left panel of Fig.~\ref{fig:fig4}, where we plot the $y$ and $z$ components of the quaternion error states ($\delta q_y$ and $\delta q_z$). As the information about the horizon determines the orientation of the satellite with respect to the orbital plane, the error of the quaternion component that describes the rotation within this plane ($\delta q_O$) does not increase during the 'night' periods. $\delta q_O$ can be produced as a linear combination of $\delta q_y$ and $\delta q_z$ in our example. The right panel of Fig.~\ref{fig:fig4} shows a short time period around a 'night' to 'day' transition and how the attitude information is immediately recovered once the Sun is visible again.

We also investigated the statistical behavior of the measurement errors. To do so we initialized our simulation with the same parameters except for the direction of angular momentum vector, which we picked randomly. By starting the simulation from several different initial conditions we collected statistical data about the first 'day' and 'night' phases.

\begin{figure}
\begin{center}
    \resizebox{10cm}{!}{\includegraphics{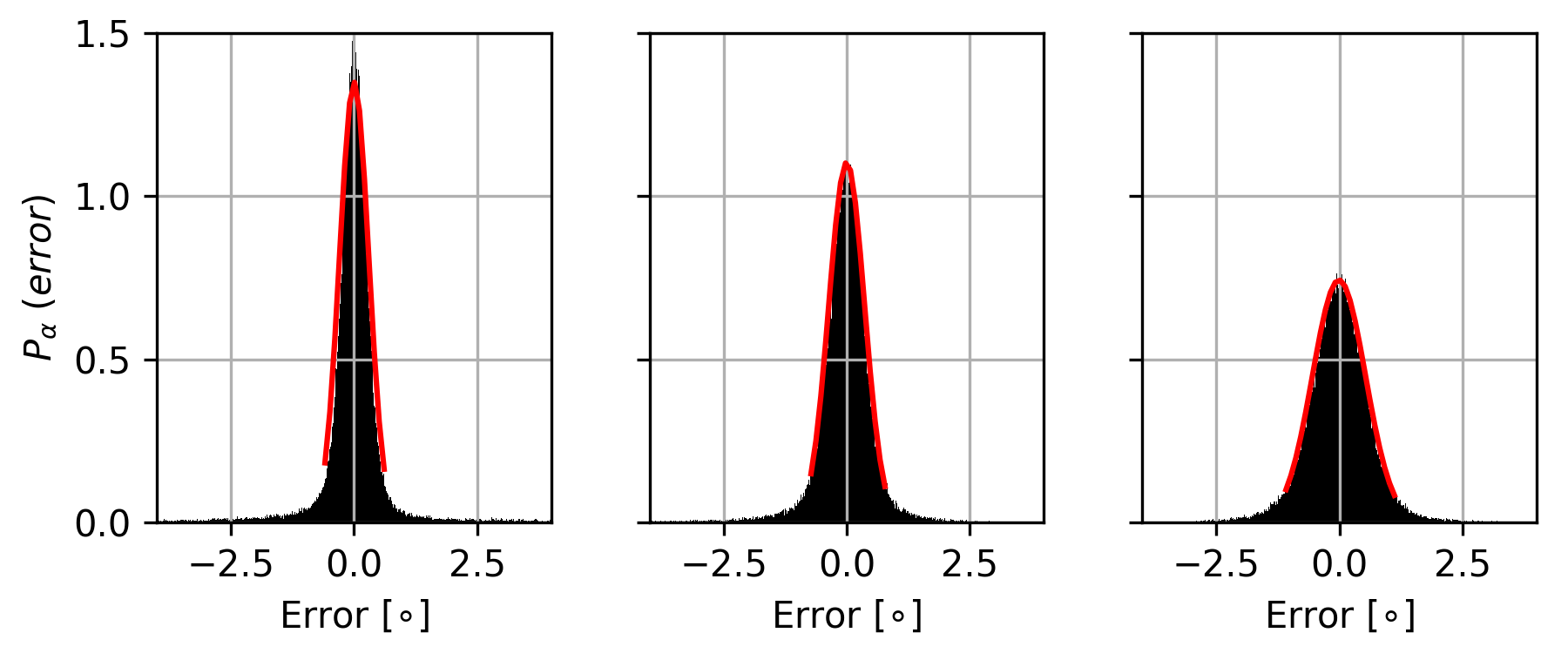}}
    \caption{Probability distributions of the right ascension's measurement error, i.e. the difference between the real and the recovered values, during 'day'. The different panels show the cases with low (left), standard (middle), and the high (right) gyroscope error. The red curves represent Gaussian fits.}
    \label{fig:fig5}
    \end{center}
\end{figure}

\begin{figure}
\begin{center}
    \resizebox{11.5cm}{!}{\includegraphics{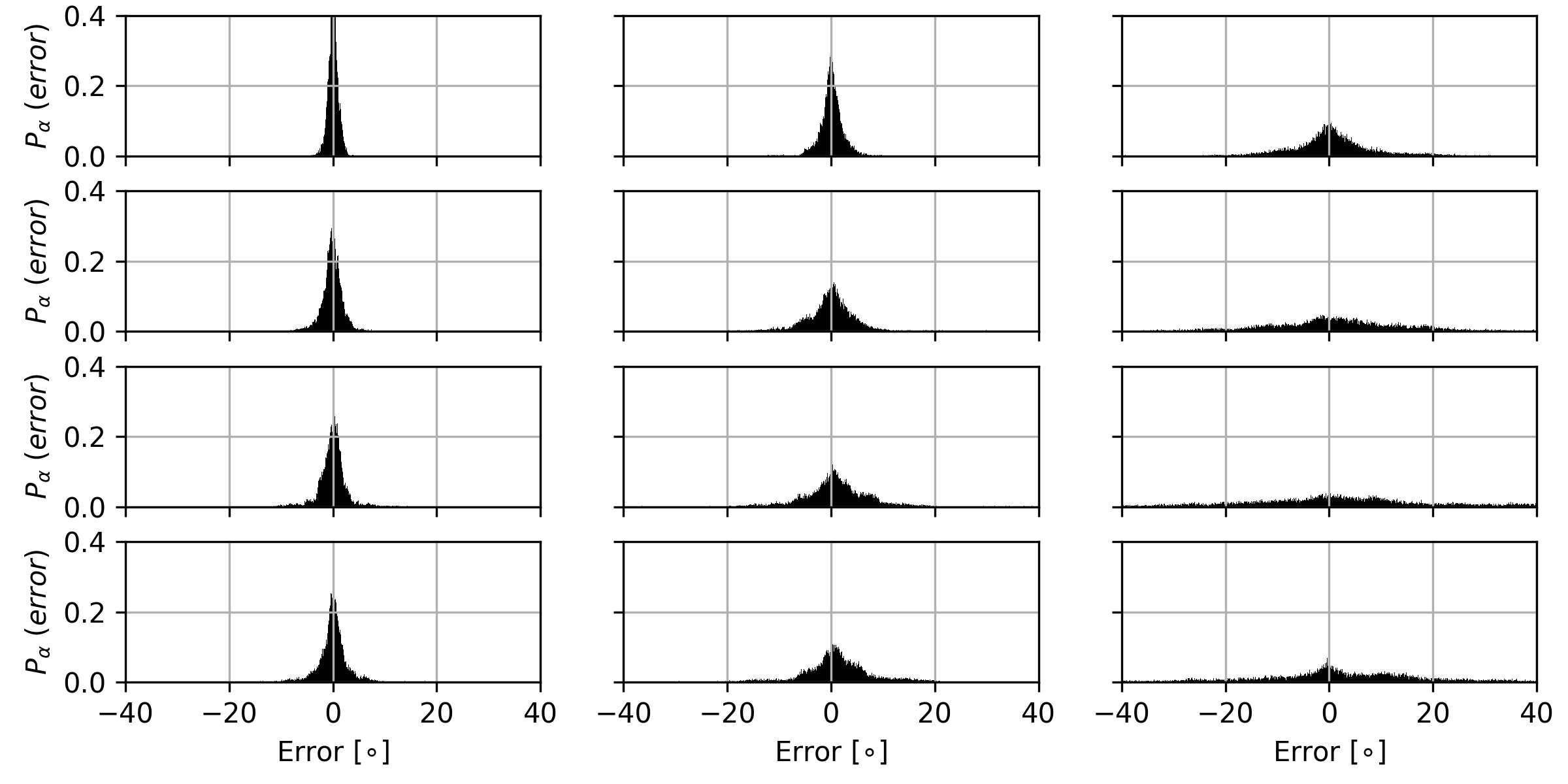}}
    \caption{Probability distributions of the right ascension's measurement error during 'night'. The left, middle and right columns correspond to the cases with low, standard and high gyroscope error, respectively, while the different rows represent different equal-length time segments with the top row being the first and the bottom row the last quarter of the 'night' phase.}
    \label{fig:fig6}
    \end{center}
\end{figure}

Figure \ref{fig:fig5} shows the distribution of the right ascension's measurement error for the 'day' case with different gyroscope precisions (the other two attitude parameters have similar error distributions). The results show that the recovery has a 1$\sigma$ error of $\sim22'$ in our standard case, while this error is $\sim18'$ and $\sim32'$ in the low- and high-error cases, respectively. This is an improvement on the $\sim40'$ error of the MLX infrasensor's point source detection accuracy \citep{kapas2021}, which shows the power of the MEKF method.

In Figure~\ref{fig:fig6} we show the same errors for different parts of the 'night' phase. We divided the 'night' period to four equal-length segments to investigate the evolution of the errors and to avoid creating statistics from time periods with qualitatively different behavior. We see that the distribution of errors gets smeared as time progresses, and also with larger gyroscope errors. In the last segment of the high-error case the distribution is completely smeared so that not much information is retained about the real attitude. This is in accordance with the results of \cite{baroni2018}.

\section{Summary}
\label{sec:conclusions}
In the present paper we described a simulation model for testing our new concept aimed at determining the attitude of nano-satellites. The attitude was represented by unit quaternions and a MEKF approach was applied to estimate the most probable state (attitude) of the system. In our model the prediction step of the Kalman filter utilizes gyroscope measurements while its measurement step is based on infrasensor measurements and GPS location information which provide the direction of the Sun and the nadir in the satellite and in the J2000 reference frames, respectively.

The results of our simulations show that an attitude accuracy of $~22'$ is achievable using combined measurements of the infrasensors and MEMS gyroscopes having a conservative drift. This is an improvement on the accuracy of point source detection with the MLX infrasensors \citep[$\sim40'$, see][]{kapas2021}. This accuracy is gradually lost when the Sun is occulted by the Earth whereupon it can reach values of $\sim15-25^\circ$. The attitude information, however, is recovered within a short time once the Sun is observed again.

During the actual mission the satellites will not have infrasensors on all of their six sides and hence not being able to observe the Sun will be more regular. However, these time periods will be relatively short and the gyroscope drift is expected to be manageable during these intervals.

In this work we simulated a satellite on LEO with an inclination of $60^\circ$. This is a reasonable choice for a particle detector experiment like a GRB detector because on this orbit the satellite evades high latitudes with increased noise contamination from the polar regions but is able to cover a large area of the sky. However, on such an orbit the illumination conditions may change substantially on the timescales of a few months due to the motion of the Earth around the Sun, as well as due to the orbital precession caused by $J_2$. However, we consider our simulations to represent the average conditions on such an orbit sufficiently well.

The attitude determination method described in this paper is planned to be used in the CAMELOT mission where the attitude data will also serve as additional information for localizing gamma-ray bursts besides triangulation. An in-orbit demonstration of our
experiment is planned to be scheduled for the end of 2022.

\begin{acknowledgement}
The authors would like to thank the support of the Hungarian Academy of Sciences via the grant KEP-7/2018, providing the financial background of our experiments. This research has been supported by the European Union, co-financed by the European Social Fund (Research and development activities at the E\"otv\"os Lor\'and University's Campus in Szombathely, EFOP-3.6.1-16-2016-00023). We also thank the support of the GINOP-2.3.2-15-2016-00033 project which is funded by the Hungarian National Research, Development and Innovation Fund together with the European Union. 
\end{acknowledgement}

\noindent\small{\textbf{Conflict of interest} The authors declare that the research was conducted in the absence of any commercial or financial relationships that could be construed as a potential conflict of interest.}

\begin{appendices}
\section{Time and measurement update with the Kalman filter}
\label{sec:appendix}

Here we describe the discretized time update and measurement update steps of the Kalman filter. From now on the superscript '--' denotes propagated states before the measurement update, while the superscript '+' denotes states after the measurement update.

\subsection{Time update}
The estimated values of the attitude quaternion and the bias vector after a $\Delta t$ time-step can be calculated using the following equations \citep{crassidis2012}:
\begin{equation}
    \hat{\mathbf{q}}_{k+1}^- = \mathbf{\Theta}(\hat{\bm{\omega}})\hat{\mathbf{q}}_k^{+} ,
\end{equation}
\begin{equation}
    \hat{\bm{\beta}}_{k+1}^- = \hat{\bm{\beta}}_k^+ ,
\end{equation}
where $\hat{\bm{\omega}}_k^+ = \bm{\omega}_m - \hat{\bm{\beta}}_k^+$ and
\begin{equation}
    \mathbf{\Theta}(\hat{\bm{\omega}}_k^+) = \begin{bmatrix}
            \cos (\frac{1}{2}||\hat{\bm{\omega}}_k^+||\Delta t) \mathbf{I}_{3\times 3} - \mathbf{S}(\hat{\psi}_k^+) &
            \hat{\psi}_k^+ \\
            -\left(\hat{\psi}_k^+\right)^{\mathrm{T}} &
            \cos (\frac{1}{2}||\hat{\bm{\omega}}_k^+||\Delta t)
            \end{bmatrix} ,
\end{equation}
\begin{equation}
    \hat{\psi}_k^+ = \frac{\sin(\frac{1}{2}||\hat{\bm{\omega}}_k^+||\Delta t)\, \hat{\bm{\omega}}_k^+}{||\hat{\bm{\omega}}_k^+||} .
\end{equation}
The covariance matrices are propagated using
\begin{equation}
    \mathbf{P}_{k+1}^- = \mathbf{\Phi}_k \mathbf{P}_k^+ \mathbf{\Phi}_k^{\mathrm{T}} + \mathbf{G}_k \mathbf{Q}_k \mathbf{G}_k^{\mathrm{T}} ,
\end{equation}
with $\mathbf{\Phi}$ being the state transition matrix:
\begin{equation}
    \mathbf{\Phi}_{k} = \begin{bmatrix}
                            \mathbf{\Phi}_{11} & \mathbf{\Phi}_{12}\\
                            \mathbf{\Phi}_{21} & \mathbf{\Phi}_{22}
                        \end{bmatrix} ,
\end{equation}
where
\begin{equation*}
    \mathbf{\Phi}_{11} = \mathbf{I}_{3\times 3} - \mathbf{S}(\hat{\bm{\omega}}_k^+)\,\frac{\sin (||\hat{\bm{\omega}}_k^+||\Delta t)}{||\hat{\bm{\omega}}_k^+||} + \mathbf{S}(\hat{\bm{\omega}}_k^+)^2\,\frac{[1-\cos (||\hat{\bm{\omega}}_k^+||\Delta t)]}{||\hat{\bm{\omega}}_k^+||^2} ,
\end{equation*}
\begin{equation*}
    \mathbf{\Phi}_{12} = \mathbf{S}(\hat{\bm{\omega}}_k^+)\,\frac{[1-\cos (||\hat{\bm{\omega}}_k^+||\Delta t)]}{||\hat{\bm{\omega}}_k^+||^2} - \mathbf{I}_{3\times 3}\Delta t - \mathbf{S}(\hat{\bm{\omega}}_k^+)^2\,\frac{[||\hat{\bm{\omega}}_k^+||\Delta t - \sin (||\hat{\bm{\omega}}_k^+||\Delta t)]}{||\hat{\bm{\omega}}_k^+||^3} ,
\end{equation*}
\begin{equation}
    \mathbf{\Phi}_{21} = \mathbf{0}_{3\times 3} , \qquad
    \mathbf{\Phi}_{22} = \mathbf{I}_{3\times 3} .
\end{equation}
The $\mathbf{Q}$ and $\mathbf{G}$ matrices determining the process noise matrix are given by
\begin{equation}
    \mathbf{Q}_k = \begin{bmatrix}
                        \left(\sigma_{\omega}^2\Delta t + 1/3\ \sigma_{\beta}^2\Delta t^3\right)\mathbf{I}_{3\times 3} & 
                        -\left(1/2\ \sigma_{\beta}^2\Delta t^2\right) \mathbf{I}_{3\times 3}\\
                        -\left(1/2\ \sigma_{\beta}^2\Delta t^2\right) \mathbf{I}_{3\times 3} &
                        \left(\sigma_{\beta}^2\Delta t\right) \mathbf{I}_{3\times 3}
                    \end{bmatrix} ,
\end{equation}
\begin{equation}
    \mathbf{G}_k = \begin{bmatrix}
                        -\mathbf{I}_{3\times 3} & \mathbf{0}_{3\times 3}\\
                        \mathbf{0}_{3\times 3} & \mathbf{I}_{3\times 3}
                    \end{bmatrix} .
\end{equation}

\subsection{Measurement update}
In the measurement update step the MEKF first estimates the quaternion error state $\delta\mathbf{q}$ using the sensitivity matrix determined by Eq. (\ref{eq:dq}) and then updates the attitude utilizing Eq. (\ref{eq:dq+-}).

Supposing there are $n$ vectors measured by the satellite, the quaternion error state and the bias vector error can be obtained using the following formula:
\begin{equation}
    \begin{bmatrix}
    \delta \mathbf{q}_k^3\\
    \delta \bm{\beta}_k
    \end{bmatrix}
    = \mathbf{K}_k
    \left(
    \begin{bmatrix}
    \mathbf{r}_{1,k}^s\\
    \mathbf{r}_{2,k}^s\\
    \vdots\\
    \mathbf{r}_{n,k}^s
    \end{bmatrix}
    -
    \begin{bmatrix}
    \hat{\mathbf{r}}_{1,k}^s\\
    \hat{\mathbf{r}}_{2,k}^s\\
    \vdots\\
    \hat{\mathbf{r}}_{n,k}^s\\
    \end{bmatrix}
    \right) ,
\end{equation}
where $\mathbf{r}_{i}^s$ denotes a vector measured by the satellite, while $\hat{\mathbf{r}}_{i}^s = \mathbf{A}(\hat{\mathbf{q}}^-) \mathbf{r}_{i}^i$ is the predicted value of that vector. $\mathbf{K}_k$ is the Kalman gain defined the usual way:
\begin{equation}
    \mathbf{K}_k = \mathbf{P}_k^- \mathbf{H}_k^T \left(  \mathbf{H}_k \mathbf{P}_k^- \mathbf{H}_k^T + \mathbf{R}_k \right)^{-1} ,
\end{equation}
with $\mathbf{H}_k$ being the sensitivity matrix:
\begin{equation}
    \mathbf{H}_k = 
    \begin{bmatrix}
    2 \mathbf{S}(\hat{\mathbf{r}}_{1,k}^s) & 0_{3 \times 3}\\
    2 \mathbf{S}(\hat{\mathbf{r}}_{2,k}^s) & 0_{3 \times 3}\\
    \vdots & \vdots\\
    2 \mathbf{S}(\hat{\mathbf{r}}_{n,k}^s) & 0_{3 \times 3}\\
    \end{bmatrix} ,
\end{equation}
and $\mathbf{R}_k$ the measurement covariance matrix:
\begin{equation}
    \mathbf{R}_k = 
    \mathrm{diag}[\sigma_{\mathbf{r}_1}^2 \mathbf{I}_{3 \times 3}, \sigma_{\mathbf{r}_2}^2 \mathbf{I}_{3 \times 3}, \hdots, \sigma_{\mathbf{r}_n}^2 \mathbf{I}_{3 \times 3}] .
\end{equation}
The quaternion state, the bias vector and the covariance matrix are then updated by
\begin{equation}
     \hat{\mathbf{q}}_k^{+} = \delta\mathbf{q}_k \otimes \hat{\mathbf{q}}_{k}^- ,
\end{equation}
\begin{equation}
    \bm{\beta}_k^+ = \bm{\beta}_k^- + \delta \bm{\beta}_k ,
\end{equation}
\begin{equation}
    \mathbf{P}_k^+ = \left( \mathbf{I} - \mathbf{K}_k \mathbf{H}_k \right) \mathbf{P}_k^- ,
\end{equation}
where the quaternion error state is obtained from its imaginary part using the normalization constraint:
\begin{equation}
    \delta \mathbf{q} = \left[ \delta {\mathbf{q}_k^{3}}^T , \sqrt{1 - \delta {\mathbf{q}_k^{3}}^T \cdot\\ \delta \mathbf{q}_k^3} \right]^T .
\end{equation}

In our setup the number of measured vectors is $n=2$ when the Sun and the horizon is visible at the same time, while it is $n=1$ when the Sun is occulted by the Earth.

\end{appendices}

{}

\end{document}


\pagenumbering{gobble}

\textbf{Description for supplementary material:} This figure shows the evolution of a rigid body when the angular momentum vector initially points towards the close vicinity of the axis corresponding to the intermediate moment of inertia. The blue, green and red arrows correspond to the x, y and z axis of the principal coordinate system, respectively, with increasing moments of inertia. The black and brown arrows are the angular velocity and angular momentum vectors, respectively. In this case there are no external torques acting on the system, hence the angular momentum remains constant.